\documentclass[%
 reprint,
 amsmath,amssymb,
 aps,
]{revtex4-2}

\usepackage{graphicx}
\usepackage{dcolumn}
\usepackage{bm}
\usepackage{textcomp}
\usepackage{braket}
\usepackage{epstopdf}
\usepackage{xcolor}
\usepackage{booktabs}
\usepackage{hyperref}

\begin{document}

\preprint{APS/123-QED}

\title{Frequency-domain optical coherence tomography with undetected mid-infrared photons}

\author{Aron Vanselow$^1$, Paul Kaufmann$^1$, Ivan Zorin$^2$, Bettina Heise$^2$, Helen M. Chrzanowski$^1$, and Sven Ramelow$^{1,3,*}$}
 \affiliation{%
  $^1$Institut f\"ur Physik,\\
  Humboldt-Universit\"at zu Berlin, Newtonstr. 15, 12489 Berlin, Germany
  \\$^2$Research Center for Materials Characterization and Non-Destructive Testing GmbH,\\
  Science Park 2, Altenberger Str. 69, 4040 Linz, Austria
  \\$^3$IRIS Adlershof,\\
  Humboldt-Universit\"at zu Berlin, Zum Gro{\ss}en Windkanal 6, 12489 Berlin, Germany
  \\$^*$Corresponding author: sven.ramelow@physik.hu-berlin.de
 }%




\date{\today}

\begin{abstract}
Mid-infrared light scatters much less than shorter wavelengths, allowing greatly enhanced penetration depths for optical imaging techniques such as optical coherence tomography (OCT). However, both detection and broadband sources in the mid-IR are technologically challenging. Interfering entangled photons in a nonlinear interferometer enables sensing with undetected photons making mid-IR sources and detectors obsolete.
Here we implement mid-infrared frequency-domain OCT based on ultra-broadband entangled photon pairs. We demonstrate 10 \textmu m axial and 20 \textmu m lateral resolution 2D and 3D imaging of strongly scattering ceramic and paint samples. Together with $10^6$ times less noise scaled for the same amount of probe light and also vastly reduced footprint and technical complexity this technique can outperform conventional approaches with classical mid-IR light.
\end{abstract}

\setlength{\parskip}{0pt}
\maketitle


Optical coherence tomography (OCT) \cite{youngquist86,Huang91} is a powerful imaging technique yielding detailed morphological information about sub-surface structures. It distinguishes itself favourably against competing techniques, notably by its factor 10 to 100 resolution enhancement over ultrasound, and its significantly enhanced sample penetration when compared to confocal microscopy \cite{Tuchin13}. Since its inception, OCT has become a widely used tool for three-dimensional scans of biological tissue \cite{zysk07}, most notably for imaging cross-sections of the human retina in glaucoma and macular degeneration diagnostics \cite{Sakata09} as well as finding application in cardiology \cite{Jorge15} or dermatology \cite{Olsen18}.

So far, all commercialized applications of OCT have exploited visible or near-infrared (near-IR) light, owing to the transmission properties of imaged samples and availability of suitable sources and detectors. However, in strongly scattering media, such as ceramics, paints, and micro-porous or micro-fibrous materials, visible and near-IR light has a very shallow penetration depth, rendering OCT largely ineffective. This is especially relevant for applications in non-destructive testing \cite{stifter07} where these materials play a major role. 

This challenge can be addressed by using much less scattering mid-infrared (mid-IR) light \cite{Colley07,rongsu14,israelsen18}. An important application for mid-IR OCT lies in alumina-based ceramics in microfluidics. There the well-defined size and high density of the pores make these ceramics useful for drug testing \cite{liu09} as well as DNA detection and separation \cite{vlassiouk04,kim08}, while the strong scattering from the pores is considerably reduced for mid-IR illumination \cite{rongsu14,israelsen18,zorin18}. Real-time OCT measurements in microfluidics may also be used for studying liquid mixing dynamics \cite{xi04}. Other examples for applications of non-destructive testing with mid-IR OCT include coating analysis in pharmaceuticals \cite{koller11} and art analysis \cite{zorin18}. 

Despite its broad applicability, there remains no commercial system for mid-IR OCT. This reflects the fact that all existing implementations were so far largely technologically constrained, either by their demand for expensive and complex light sources, or their dependence on inferior detection technologies that, in comparison with their visible and near IR counterparts, combine low efficiency with large noise. At its root this limitation is unfortunately fundamental: when using semiconductors for detectors and light-sources one has to - while competing with the thermal noise - match the band-gap with the photon-energy, which for the mid-IR is almost an order of magnitude lower than that of the visible to near-IR.

Consequently, mid-IR OCT was only first demonstrated in 2007 \cite{Colley07}. It utilised a pulsed quantum cascade laser (QCL) source and liquid-nitrogen cooled detector resulting in a very long A-scan time of about 30 minutes compared to microseconds for commercial systems in the visible. More recently, a comparatively bright super-continuum light source was combined with a low-cost pyro-electric thermal detector to implement OCT in the mid-IR, yielding approximately 50 \textmu m axial resolution, and a  signal-to-noise ratio (SNR) normalized to the A-scan time of 87 dB/s \cite{zorin18}. Techniques circumventing the need for mid-IR detectors have also recently emerged. These techniques exploit broadband up-conversion \cite{Hu:12}, allowing detection with standard silicon detectors \cite{israelsen18}. Combined with a super-continuum source, they achieved an axial resolution of 9 \textmu m and 85 dB/s sensitivity normalized for A-scan times of 3 ms. In both works, as much as 20 mW of average optical power are tightly focused onto the samples, resulting in average intensities in excess of 1 kW/cm$^2$.

Here, we present a fundamentally different approach to mid-IR OCT based on nonlinear interferometry exploiting broadband, entangled photon pairs, thereby circumventing the need for mid-IR lasers or detectors. Our technique thus enjoys significantly enhanced experimental ease and cost-effectiveness, employing only room-temperature off-the-shelf detection technology and a standard, visible continuous-wave (CW) laser. Moreover, we find a 6 orders-of-magnitude improvement for the integration-time normalized SNR per power on the sample due to the inherently low spectral noise of our method which we find is primarily only affected by shot-noise. Thus, with maximum mid-IR power of only $\approx 90$ pW on the sample we experimentally demonstrate a normalized SNR of 66 dB/s with 10 \textmu m axial resolution. Further, we demonstrate the method's relevance for real-world samples, with 2D and 3D imaging of thick alumina ceramic structures and paint layers, which are inaccessible for commercial OCT systems.

OCT exploits the interference of low coherence light to reveal depth information from within a scattering or layered structure. It is typically implemented as a Michelson interferometer, where light in the reference arm interferes with probe light in the sample arm. In time-domain OCT (TD-OCT), scanning the reference mirror results in an interferogram that reveals the depth information, while in frequency-domain OCT (FD-OCT) the reference mirror remains fixed and spectral interference is measured via a spectrometer. A subsequent Fourier transform then yields the desired depth information. Owing to the absence of a scanning mirror, the frequency domain variant offers significant advantages in both speed and mechanical stability. While TD-OCT can only achieve A-scan rates of few kHz, FD-OCT can typically achieve rates of up to 300 kHz, limited primarily by the acquisition rate of the spectrometer \cite{wieser15}. For a fixed integration time, FD-OCT also has a fundamental advantage in sensitivity (20 to 30 dB) \cite{choma03} over TD-OCT. As a consequence of these advantages, FD-OCT has almost fully supplanted TD-OCT \cite{wieser15}. 

To realise OCT in the mid-IR, while requiring neither mid-IR detection or laser source capabilities, one can exploit nonlinear interferometry with pairs of photons generated via spontaneous parametric down-conversion (SPDC) \cite{Chekhova:2016hw}. Recent works have considered this approach so far only in the context of time-domain OCT \cite{valles18,paterova18}. Also, these initial demonstrations proved comparatively slow and insensitive, realising only relatively low and practically insufficient axial resolutions of 500 \textmu m at 1550 nm \cite{valles18} and 105 \textmu m in the mid-IR \cite{paterova18}. Recent theoretical works have considered time-domain OCT without entanglement \cite{RojasSantana:2020tk} and experimental measurements in the high-gain regime \cite{Machado:2020tu}. In addition to OCT, the technique of using nonlinear interferometers for "sensing with undetected photons" has seen use for other modalities of sensing, such as imaging \cite{Lemos14}, spectroscopy \cite{Kalashnikov:2016cl}, and refractometry \cite{paterova18}.  

\begin{figure*} [htbp]
\centering
\includegraphics[width=0.8\linewidth]{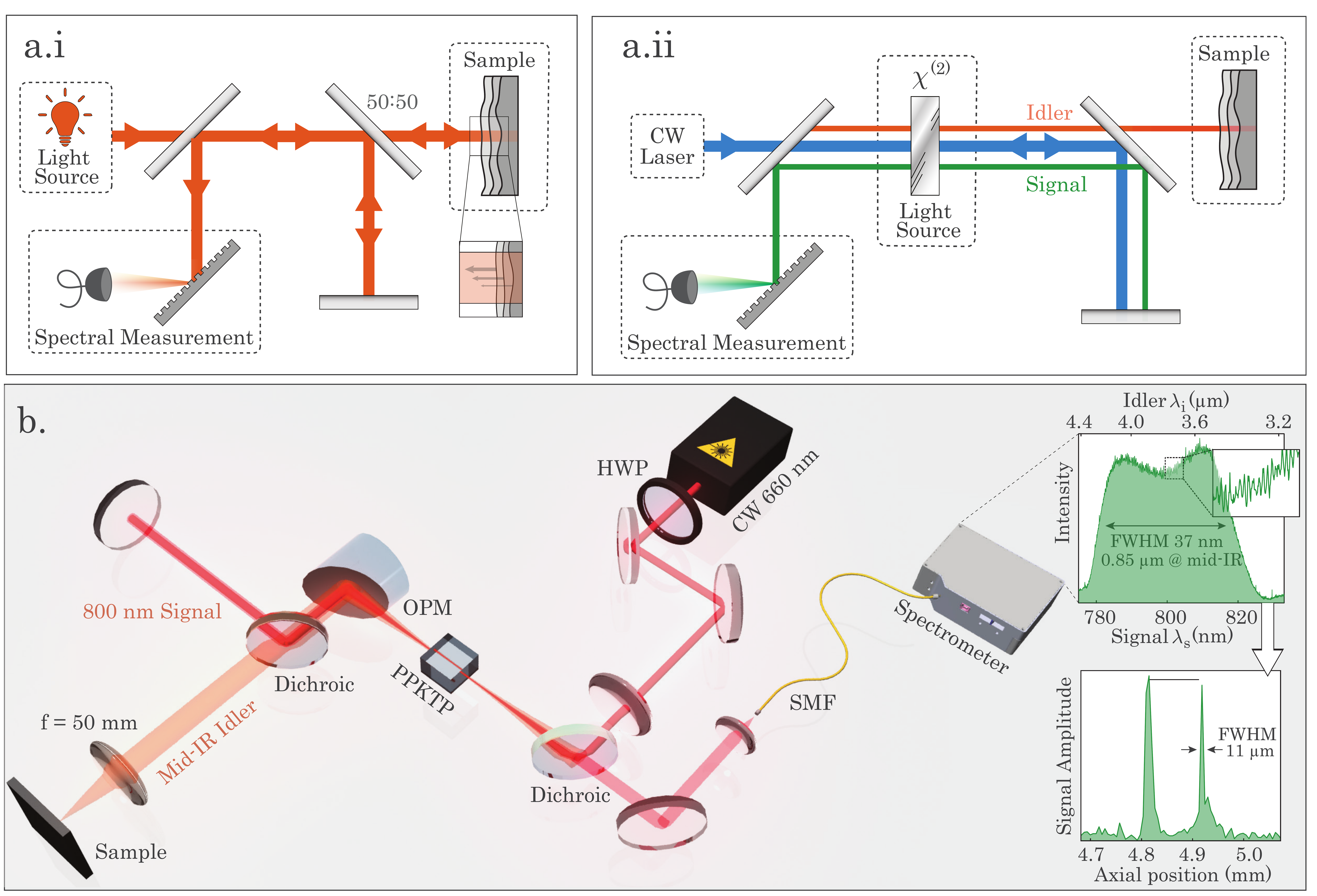}
\caption{{\bf a.} Working principle underlying {\bf i.} classical FD-OCT and {\bf ii.} FD-OCT with undetected photons. {\bf b.} {\it Experimental setup: } The nonlinear interferometer is formed by double-passing a nonlinear crystal in a Michelson geometry. Pump light from a 660 nm diode-pumped solid state laser is focused into a 2 mm long ppKTP crystal phase-matched for type-0 SPDC. It produces pairs of signal and idler photons at central wavelengths of 798 nm and 3.8 \textmu m respectively. The pump light and resulting signal and idler photon are collimated via a 90$^\circ$ off-axis parabolic mirror (OAP). A long-pass dichroic mirror reflects the pump and signal fields whilst transmitting the idler.  The idler photons are focused onto the sample, forming the end mirror of the interferometer arm. The signal and pump are reflected back via a silvered mirror. All fields propagate back to the nonlinear crystal and thereby close the interferometer. Finally, the signal light is detected using a commercial, uncooled spectrometer. The spectrometer has a bin width of around 0.04 nm and a resolution (FWHM) of around 0.11 nm (effectively 2.5 nm at 3.8 \textmu m). The inset shows typical raw spectrum, here taken for a microscope slide. After linearistion, dispersion-compensation and Fourier-transform and accounting for the double-pass and group-index it results in the A-scan shown in the second inset, yielding the expected slide thickness of 100 \textmu m and resolution of the system of 10 \textmu m.}
\label{fig:setup}
\end{figure*}

The concept and its experimental realisation is presented in fig. \ref{fig:setup}. A 660 nm CW laser pumps two SPDC processes in a nonlinear crystal --- first in the forward and second in the backward direction through the same crystal. This double-pass configuration entails significant experimental ease for implementing the nonlinear interferometer. The nonlinear crystal is purpose engineered to generate very broadband and widely non-degenerate photon pairs \cite{vanselow193} with signal photons in the near-IR for detection and idler photons in the mid-IR for probing the sample. Between the first and the second pass through the crystal, the pump and signal fields are separated from the idler via a dichroic mirror, their respective paths forming the two arms of an interferometer. The idler probes and is back-reflected from a sample (which potentially constitutes several reflectors with individual displacements and reflectivities). The pump and signal are reflected back via a silvered mirror, with the interferometer arm length approximately matched to that traversed by the idler. All fields then propagate back and are realigned back into the crystal. Consider the simple scenario where the sample is a single reflector with an amplitude reflectivity, $r_1$. Considering only a single spectral mode of the biphoton, with a signal wavelength $\omega_s$ (and accordingly an idler wavelength $\omega_i = \omega_p -\omega_s$), the state vector after the second crystal is given by 
\begin{align}
\ket{\psi}=&\frac{1}{\sqrt{2}} s(\omega _s) ( r_1 \ket{1_s , 1_i} + \nonumber  \\ & + \sqrt{1-r_1^2}\ket{1_s , 0_i} + e^{i\Delta \phi}\ket{1_s , 1_i}  ),
\end{align}
where we have assumed the first and second SPDC processes are identical, and $s(\omega _s)$ is the amplitude associated with the mode $\omega_s$ and specified by the phase-matching properties of the crystal. The interference is determined by the collective phase mismatch acquired between all the three fields ($\Delta \phi =\phi_p-\phi_s-\phi_i$) between the centre of the first and second crystal \cite{Chekhova:2016hw}. An intensity measurement of the resulting signal field gives
\begin{align}
    I = |s(\omega _s)|^2  (1+ r_1 \cos{(\Delta \phi)}). 
\end{align}
Crucially, the phase acquired by the idler photon will influence the final intensity of the measured signal photons. To illuminate its relevance for OCT, we recast $\Delta \phi$ in terms of the group delay $\tau=-\frac{\partial\phi}{\partial\omega}$ and obtain
\begin{align}
        \Delta \phi=&\phi_{p}-\phi_{s_0}-\phi_{i_0}-\dfrac{\partial\phi_s}{\partial\omega_s}|_{\omega_{s_0}}\Delta\omega_s-\dfrac{\partial\phi_i}{\partial\omega_i}|_{\omega_{i_0}}\Delta\omega_i \nonumber
    \\=&D-\Delta\omega_s\Delta\tau 
\end{align}
where the last equality arises due to energy conservation ($\omega_i=\omega_p-\omega_s$). $D$ is a constant. Thus, one obtains the following modulated spectrum
\begin{align}
I(\omega_s)=S(\omega_s)(1+r_1 \text{cos}(D-\omega_s\cdot\Delta \tau)). \label{eq:interference} 
\end{align}
$S(\omega_s)$ is the spectral envelope, which is here an approximately rectangular function obtained by dividing the measured spectrum by the spectrum obtained without interference. Applying a Fourier transform to eq.  \ref{eq:interference}, we obtain
\begin{align}
\mathcal{F}\{ I(\omega_s) \}= r_1 \tilde{S}(\tau-\Delta\tau)= r_1\tilde{S}(2\cdot (x-\Delta x)\cdot n_g / c_0).
\end{align}
Here, $\tilde{S}$ is the Fourier transform of the spectral envelope, $c_0$ is the vacuum speed of light and $n_g$ is the group index in the idler arm. The factor 2 arises as $\Delta\tau$ is the round-trip group delay.  If the sample contains several partially reflecting reflectors, the interference term becomes a sum of interference terms, each with the respective spectral phase and weighted by their respective amplitude. The Fourier transform is the sum of the individual Fourier transforms, weighted in the same manner, thus revealing the reflectances and positions of all reflectors. This is entirely analogous to how FD-OCT works, albeit applied to nonlinear interferometry.

Accordingly our OCT-measurements are evaluated by Fourier transforming the measured and re-sampled spectra, described in detail in the supplementary information. Dispersion compensation is implemented both physically (by placing 8 mm of AR coated silicon into the mid-IR arm) and numerically in post-processing. The signal spectra are acquired using a commercial grating spectrometer spectrometer working at room temperature. In case of B- and C-scans, a sample spectra are acquired for every stage position. Each spectrum is divided by a reference spectrum without interference (with the idler field blocked). 

\begin{figure*} [htbp]
\centering
\includegraphics[width=\textwidth]{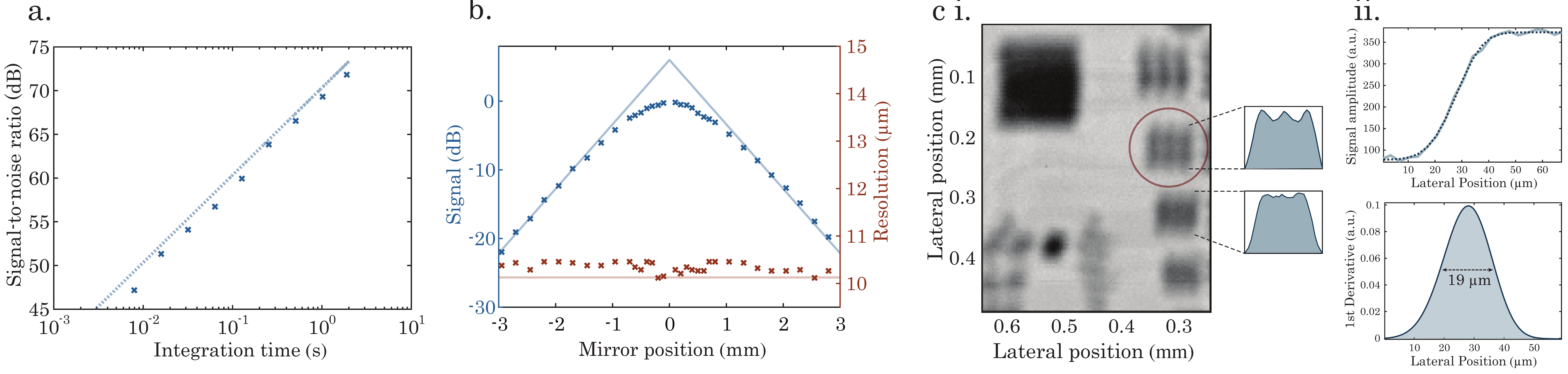}
\caption{{\bf a.} Dependency of the measured signal-to-noise ratio (SNR, blue crosses) in dB on the integration time, measured with a silver mirror in the sample arm. The blue line is the theoretical prediction. {\bf b.} Dependency of the SNR (blue) with the theory prediction using the approximation of a Lorentzian response function (blue line) and the axial resolution (red) on the mirror position with the theoretical prediction (red line). {\bf c. i.} Lateral scan of a USAF 1951 resolution target. The highlighted element 2 of group 5 is still resolved. It has a stripe width of 14 \textmu m. {\bf c. ii.} Edge response measurement. {\it Upper:} Blue: Horizontal cut through the square in the upper left corner of the resolution target in c. Dashed line: Fit with $\sqrt{\text{erf}}$ function. {\it Lower:} First derivative of the fit in the upper plot to obtain the point-spread function. The FWHM is 19 \textmu m. }
\label{fig:lateral}
\end{figure*}

Only signal wavelengths larger than 784 nm are considered, owing to the strong CO$_2$ absorption lines between the corresponding range of 4.2 -- 4.3 \textmu m in the mid-IR, which result in a decrease in visibility and strong nonlinear dispersion. Signal wavelengths larger than 822 nm are omitted due to the reduced spectral density. The FWHM of the useful signal and idler spectra is 18 THz, corresponding to 38 nm (784 nm to 822 nm) and 824 nm (3349 nm to 4173 nm), respectively.
Given this, the anticipated OCT resolution (FWHM) can be readily calculated from the available signal spectrum via the Wiener-Khinchin theorem to be 10.1 \textmu m. Characterizing the axial resolution of the setup using a silvered mirror in the sample arm, we obtained resolutions between 10.1 and 10.5 \textmu m for mirror positions between -3 mm to 3 mm (fig. \ref{fig:lateral} b). This is in excellent agreement with the theoretical prediction. 

\begin{figure} [htbp]
\centering
\includegraphics[width=0.95\linewidth]{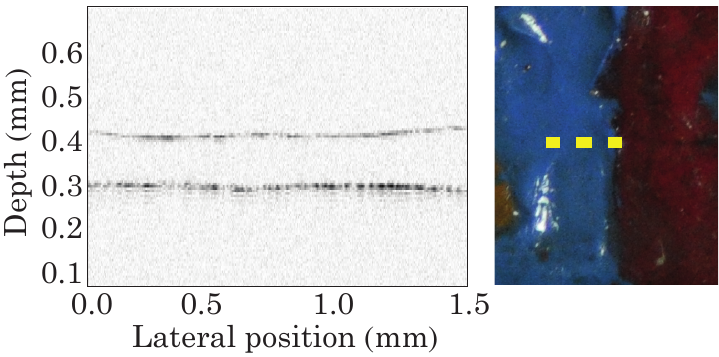}
\caption{Left: Scan of oil paint at 200 ms A-scan time. The upper reflections are from the paint layer, the lower reflections from the substrate. Right: Picture of the sample with the scanning range marked by the yellow line.}
\label{fig:Paint}
\end{figure}

The dependence of the SNR on the integration time (fig. \ref{fig:lateral} a.) was found to be approximately linear, with a SNR of 45 dB at 8 ms integration time improving to 66 dB at 1 s. We also compare the SNR with the relevant theory \cite{leitgeb03}, including an important correction that significantly improves the shot-noise limited performance when using absolute value of the Fourier-transform \cite{kalkman17} (also shown in fig. \ref{fig:lateral} a.), and find that they are in good agreement. This shows that our method comes close to the limit imposed by detection efficiency and shot-noise, and that other noise sources are practically irrelevant.
The signal roll-off, i.e. the dependence of the signal amplitude upon the mirror position, is depicted in fig. \ref{fig:lateral} b. It drops by 6 dB within 1.2 mm and by 20 dB within 2.8 mm from the zero path-length difference point. This roll-off agrees well with the theoretical expectation for the $\approx$ 350 pixels (38 nm) of the spectrometer used as raw data for the analysis. Extending this by one order-of-magnitude by using state-of-the-art 4096 pixel image sensors and high-resolution dispersion optics would lead to an operation range well beyond 15 mm.
\begin{figure} [htbp]
\centering
\includegraphics[width=0.9\linewidth]{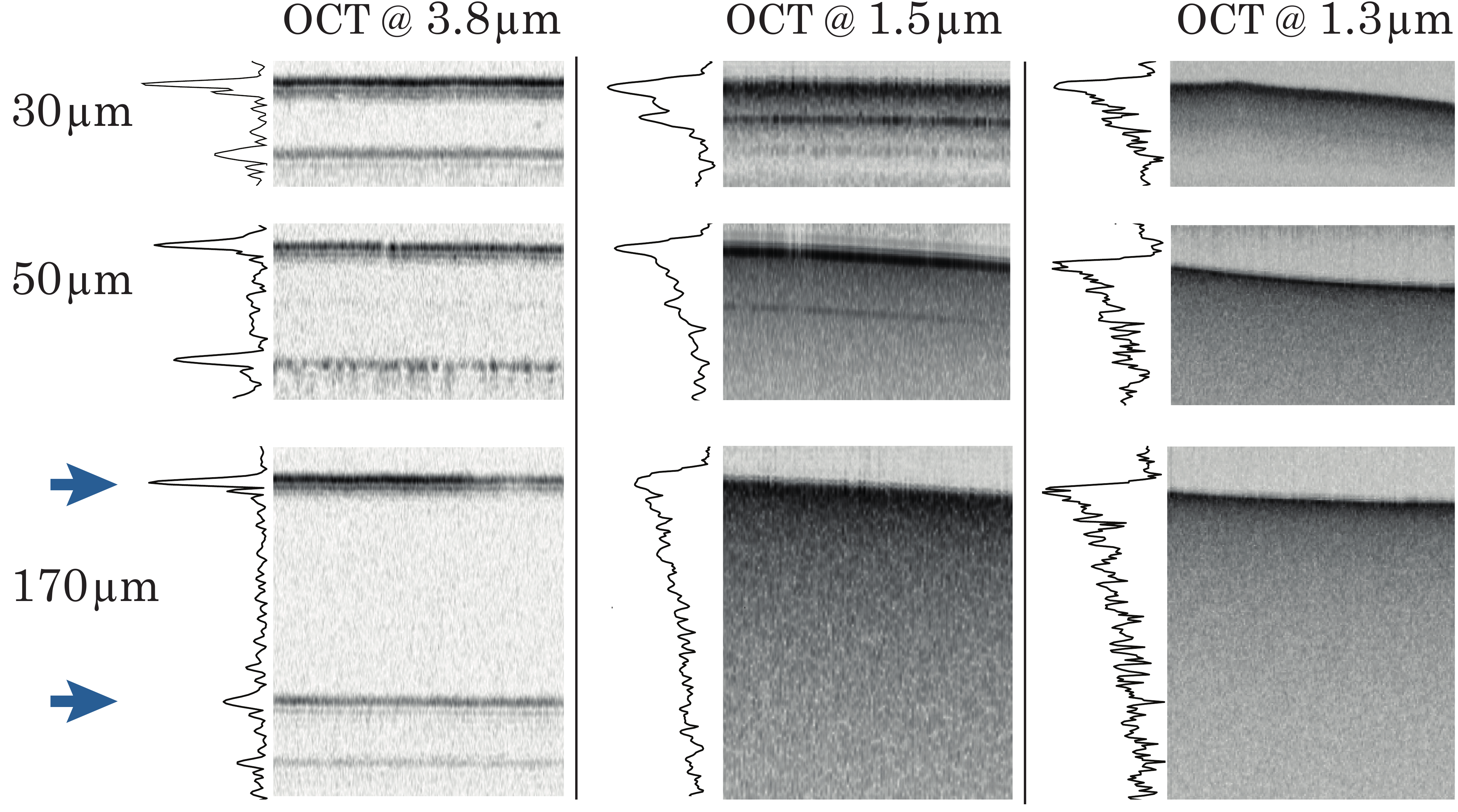}
\caption{{\bf Scan through alumina ceramic structures of varying thickness applied to a cellulose acetate foil.} Left: scan with the mid-IR setup presented in the present work. Middle: Scan with an OCT setup at 1.55 \textmu m. Right: Scan with a commercial Thorlabs Telesto OCT setup at 1.3 \textmu m. All scans have the same vertical scaling. The alumina layer lies between the top and the middle reflectors, highlighted with blue arrows for one measurement. The bottom reflector forms the boundary between the foil and air. Due to variations in the sample, the middle reflector is not visible in all measurements. With increasing alumina thickness, the 1.5 \textmu m setup fails to penetrate the sample at about 50 \textmu m alumina thickness. By comparison, the mid-IR setup can easily penetrate the ceramics at 170 \textmu m alumina thickness - the maximum alumina thickness on the sample. The 1.3 \textmu m setup struggles to penetrate even 30 \textmu m of alumina, owing to its even shorter wavelength. While scattering is clearly visible in the near-IR setups, it is effectively absent in the mid-IR setup. This becomes clear when comparing the signal strength from the front of the tape (at the top of each image) and within the tape, which are near identical for the mid-IR setup.}
\label{fig:Tape}
\end{figure}
The lateral resolution was obtained by performing a knife-edge measurement. The resulting point-spread function (PSF) has a FWHM of 19 \textmu m, see fig. \ref{fig:lateral} c. It was confirmed using a 1951 USAF resolution target. The line pairs (of element 2 in group 5) with a width of 14 \textmu m could still be resolved. Note that higher lateral resolution can readily be obtained using more magnifying optics, however at the cost of a smaller depth-of-focus. In our implementation this is 2$z_0$ = 0.4 mm due to the long wavelength and relatively tight focusing.
\begin{figure} [htbp]
\centering
\includegraphics[width=0.9\linewidth]{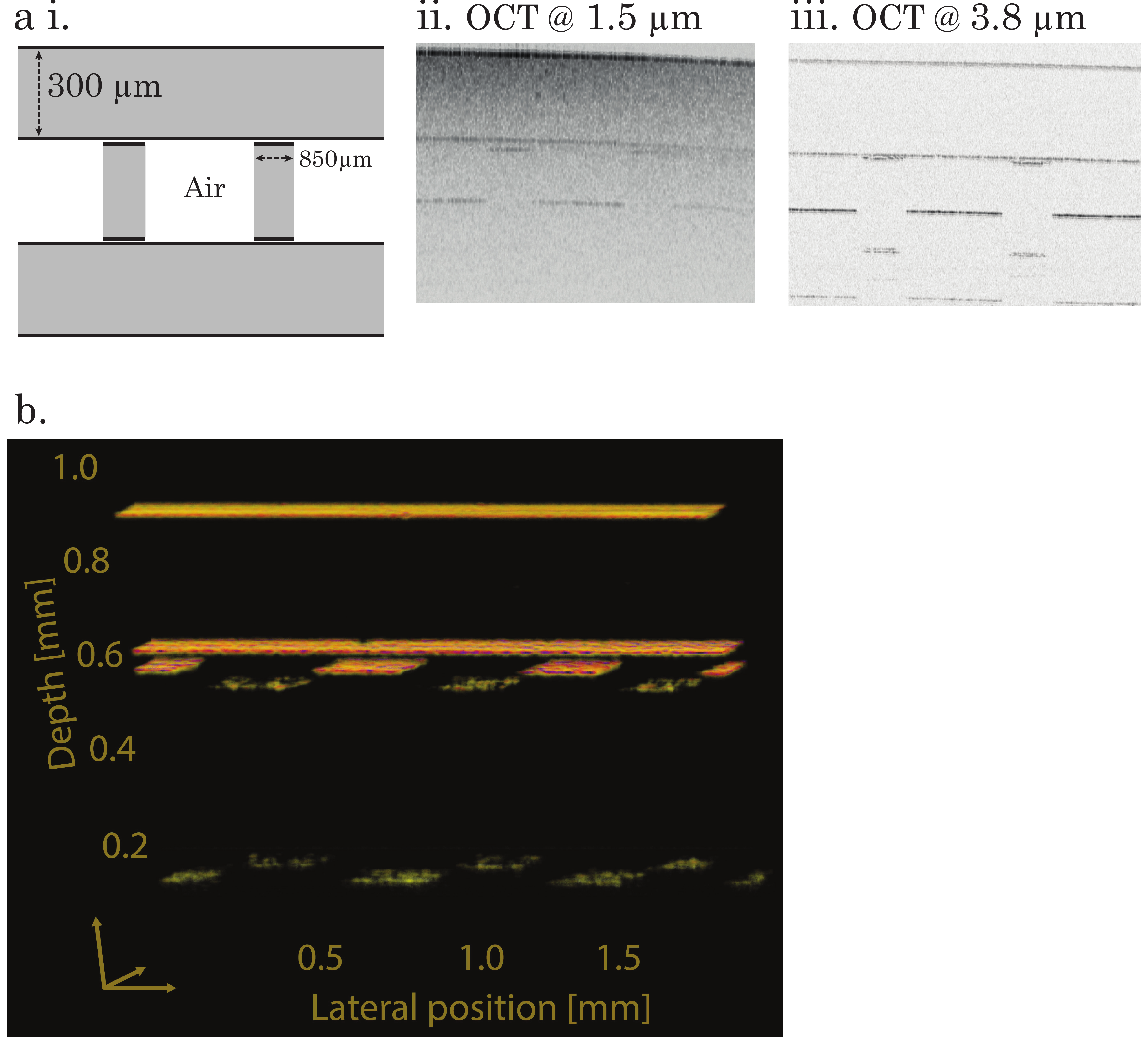}
\caption{{\bf a. i.} Schematic of the measured sample. {\bf ii.} OCT scan at 1.5 \textmu m wavelength and an A-scan integration time of 8 ms through a 900 \textmu m thick structure, consisting of three layers of alumina ceramics. {\bf iii.} Comparison scan with our presented mid-IR setup at the same 8 ms A-scan time. The scattering is greatly reduced and the penetration depth significantly increased. The air gaps between the layers are clearly resolved. The top and bottom facets of the bottom layer appear shifted up due to the difference of the group indices of air and alumina. {\bf b.} C-scan of an alumina ceramics stack with laser-written channels between the top and the bottom layer. }
\label{fig:Ceramics}
\end{figure}
To show the remarkable efficacy of the proposed method, despite the comparatively very low levels of illumination, we measured several samples of practical relevance. In fig. \ref{fig:Ceramics}, a 900 \textmu m thick stack of alumina ceramic was scanned. The stack contained small air channels suitable for applications in microfluidics. The depth information is clearly resolved for integration times as low as 8 ms. These measurements compare very favourably to equivalent measurements with an OCT setup \cite{stifter11} operating at 1.55 \textmu m, with the reduction in scattering with increasing wavelength affording much improved sample penetration. To further highlight this, the performance of the mid-IR and the 1.55 \textmu m setups for additional samples were tested, with both setups operating at 8 ms integration time. The chosen sample (provided by R. Su, Univ. of Nottingham, UK) was a cellulose acetate foil to which an alumina ceramic layer of increasing thickness was applied. As can be observed in fig. \ref{fig:Tape}, while the 1.55 \textmu m setup can only penetrate to depth of approximately 50 \textmu m, the mid-IR setup can easily penetrate the ceramic layer beyond 170 \textmu m. Additionally, similar scans were performed with a commercial 1.3 \textmu m device. Unsurprisingly, the 1.3 \textmu m results realise worse penetration depth still than those at 1.55 \textmu m. This can be attributed, at least partially, to the even shorter wavelength and consequently increased scattering.
To demonstrate the 3D-imaging capability of our setup, we additionally also performed a C-scan of an alumina ceramic stack. As shown in fig. \ref{fig:Ceramics} b our method enables a full 3D reconstruction of the sample revealing the channel structures laser written into the top facet of the bottom layer 900 \textmu m below the top surface.
Lastly, we have also demonstrated the ability to scan through oil paint (see fig. \ref{fig:Paint}). Akin to ceramic samples, scattering is reduced for longer wavelengths. The substrate and the boundary of the paint layer are clearly resolved.
In the scans of realistic samples, one can observe a significant reduction in scattering compared to the mid-IR measurements of \cite{zorin18}. We attribute this improvement to the fact that scattered idler light is dominantly scattered out of the incident spatial mode, thus revealing the which-path information and eliminating any contributions of those modes to the interference. Analogously, scattering should also be reduced in \cite{israelsen18}, as they use a single mode source and a single-mode fibre defines the detected mode.
In comparison to the time-domain approach to mid-IR OCT using quantum nonlinear interferometry \cite{paterova18}, we demonstrate one order of magnitude improvement in axial resolution. Our axial resolution is thus now comparable to typical classical OCT applications in the visible and near-infrared. The vastly improved axial resolution also reflects our engineered-for-purpose broadband SPDC process and, to a lesser extent, application of physical and numerical dispersion compensation. Additionally, the transition from TD-OCT to FD-OCT affords enhanced stability and significantly faster acquisition times. We have demonstrated the suitability of the technique for samples of practical relevance, including highly scattering samples like paint and ceramics. Our comparison measurements show that our system outperforms classical near-infrared OCT setups in penetration depth. 

Given this performance, our method is best compared with recent classical approaches to mid-IR OCT which comprise the current state-of-the-art. These approaches are based on either up-conversion \cite{israelsen18} or thermal detectors \cite{zorin18}. Techniques based on thermal detection have achieved resolutions of 50 \textmu m and a SNR of 82 dB at 300 ms integration time (with a minimum integration time of 10 ms). They remark further increases in the bandwidth to garner improvements to the resolution would pose problems in mid-IR spectrometer design \cite{zorin18}. The approach based on up-conversion reached an axial resolution of 9 \textmu m and a SNR of 60 dB at 3 ms integration time.

Whilst the resolutions reached by the classical state-of-the-art for mid-IR OCT are near-identical or significantly worse than in our case, these approaches typically achieve better signal-to-noise ratios than those presented here. However, the aforementioned classical approaches typically require tens of mW of optical power focused into the sample. By contrast, our measurements consider a maximum optical power of $\approx$ 90 pW incident on the sample - a stark difference of 8 orders of magnitude. We attribute our high per photon noise performance to the shot-noise limited performance of our system, which is out of reach so far for super-continuum based systems at this wavelength. It is also this difference that likely underlies the large performance gap that exists between mid-IR OCT technologies and their visible and near-IR counterparts, and highlights that nonlinear interferometry may offer a new pathway to close the aforementioned gap. 

Furthermore, signal-to-noise ratios achieved by the existing state-of-the-art (and beyond) are within reach for the present setup using current technology. Using the linear scaling of the SNR with the integration time for shot-noise limited detection, this can be achieved using higher pump powers, the significantly increased brightness afforded by waveguided-based SPDC, the use of a pump enhancement cavity - or a combination thereof. Moreover, additional gains to the present SNR could be obtained via an optimised pump focus, absent any increase of the pump power \cite{grice11}. The axial resolution could be improved via an further increase in the SPDC bandwidth. Sophisticated approaches to quasi-phase matching - such as chirped or interleaved longitudinal patterning - would yield increases in bandwidth resulting in wavelength scale resolution, although additional dispersion compensation would need to be required or other techniques for mitigating dispersion effects be used \cite{Bradu:15}.

In totality, these results form the groundwork for mid-IR OCT to mature as a technology, and rival its established visible and near-IR counterparts. Though potentially limited by the strong absorption of water in the infrared \cite{israelsen18}, mid-IR OCT may also prove significant for bio-medical imaging, where highly scattering samples are of particular interest \cite{petersen19}.  Extended techniques, such as Doppler OCT may prove interesting, notably in the context of microfluidics. Further, the mid-IR regime offers unique spectroscopic specificity. Harnessing that existing capacity for spectrally-resolved OCT could couple morphological information with material specificity, offering new perspectives for sensing and testing. Moreover, extending the presented OCT approach to even longer wavelengths, up to and including the THz-regime \cite{Kitahara:2020ty}, could reveal new applications.

In conclusion, here we present a fundamentally new approach to mid-IR FD-OCT based on nonlinear interferometry and spectral measurements. Harnessing 18 THz broad SPDC yields 10 \textmu m axial resolution, while we require only 90 pW mid-IR power illuminating the sample to demonstrate a normalized SNR of 66 dB/s. This constitutes a dramatic 6 orders-of-magnitude improvement for the normalized SNR per power on the sample over the state-of-the-art, which we attribute to the shot-noise limited performance of both our source and our detector. Further, we have demonstrated the method's suitability for real-world samples, with 2D and 3D imaging of thick alumina ceramic structures and paint layers, which are inaccessible for commercial OCT systems.

\section*{Funding Information}
S.R. acknowledges funding from Deutsche Forschungsgemeinschaft (RA 2842/1-1).

\section*{Acknowledgments}
S.R would like to thank Gabriella B. Lemos for fruitful discussions.

\section*{Supplementary information}

\subsection*{Setup}
The nonlinear interferometer is formed by double passing a nonlinear crystal in a Michelson geometry, and is pumped by a frequency-doubled Nd-YAG continuous-wave laser with central wavelength at 660 nm and a maximum power of 500 mW. The pump light is first focused into a 2 mm periodically-poled KTP crystal. The KTP crystal is engineered for highly non-degenerate broadband down-conversion via group velocity matching of the signal and idler fields \cite{vanselow193}. The grating period of 20.45 \textmu m phase matches a type 0 process with signal and idler central wavelengths of 798 nm and 3.8 \textmu m respectively. The transmitted pump and resulting signal and idler fields from the first pass of the ppKTP crystal are collimated via a 90$^\circ$ off-axis parabolic mirror (OAP) and incident upon a long-pass dichroic mirror that reflects the pump and signal fields whilst transmitting the idler. For the reflected signal and pump fields, a silvered mirror forms the end mirror of the interferometer. The idler beam passes through 8 mm of AR-coated silicon for physical dispersion compensation. The idler beam is focused onto the sample by an $f=50$ mm lens, with the sample forming the effective idler mirror. The sample is mounted on an X-Y-Z stage, enabling B- and C-scans. All fields propagate back into the crystal, `closing' the nonlinear interferometer. The optical path lengths traversed by the signal/pump and the idler are matched to be close to zero. The signal light emerging from the second pass through the crystal is subsequently coupled into a single-mode fibre and detected via a commercial, uncooled back-illuminated spectrometer. The spectrometer has a measured resolution (FWHM) of around 0.11 nm and its image sensor a specified 80 \% quantum efficiency at 800 nm.

In order to ascertain the mid-IR illumination power, the total power of the signal beam after the first passage of the pump through the crystal was measured using a CMOS camera for varying pump strength. After correcting for losses and detection efficiency, we obtain a pair production rate of $3.2 \times 10^6$ photon pairs per mW of pump power. This corresponds to a maximum sample illumination power of 90 pW at 3.8 \textmu m for 500 mW incident pump power.

\subsection*{Data Analysis} 

The signal spectra are measured using the aforementioned spectrometer. In case of B- and C-scans, a sample spectra is acquired for every stage position. Each spectrum is divided by a reference spectrum without interference (measured with the idler field blocked). Only signal wavelengths that are larger than 784 nm are considered because of the strong CO$_2$ absorption lines (between 4.2- 4.3 \textmu m in the mid-IR) which result in a decrease in visibility and strong nonlinear dispersion. Signal wavelengths longer than 822 nm are omitted due to the reduced spectral density leading to increased noise. The spectrum is subsequently resampled from wavelengths to frequencies, and numerical dispersion compensation is performed. In certain cases, zero-padding is applied. If necessary, a Gaussian window is applied to reduce noise introduced by the transition from the modulated to the zero-padded part of the spectrum. Lastly, a discrete fast Fourier transform is applied.

Our setup considers both physical and numerical dispersion compensation. Chromatic dispersion of second (or greater) order can cause a considerable degradation of the axial resolution. The dispersion arises from the material dispersion of optical elements. In the specific setup, GVD can be almost perfectly compensated via the addition of 8 mm of AR-coated silicon in the sample arm. Additionally, the dispersion is compensated numerically as per \cite{wojtkowski04}.

However, the process of dispersion compensation revealed the inadequacy of the existing spectrometer calibration. With the factory calibration of the spectrometer, as well as after a recalibration with a Krypton lamp, the numerical dispersion compensation exhibited a dependence on the position of the sample in the idler arm ({\it i.e.} the displacement of the interferometer). Small inaccuracies in the spectrometer calibration introduced a residual chirp of the interference frequency over the spectrum, thus reducing the resolution. This problem was also addressed in \cite{Liu10}. This was overcome by first optimising the calibration such that the resolution was almost constant for all sample positions, requiring that the calibration should no longer affect the resolution, as the numerical dispersion compensation should be independent of the sample position if the calibration is correct.



\bibliography{OCTbib}

\end{document}